\documentstyle[epsfig]{mn}

\newcommand{\bi}[1]{\mbox{\boldmath $#1$}}
\def\planck{{\it Planck }}
\def\wmap{{\it WMAP }}
\def\rh{{\bi r}}

\def\alm{a_{\ell m}}
\def\Ylm{Y_{\ell m}}
\def\Blm{B_{\ell m}}
\def\Flm{F_{\ell m}}
\def\Cl{C_{\ell}}
\def\Fl{F_{\ell}}
\def\fl{f_{\ell}}
\def\Bl{B_{\ell}}
\def\Wl{W_{\ell}}
\def\const{\rm const.}
\def\sumlm{\sum_{\ell m}}
\def\summ{\sum_{m=-\ell}^{\ell}}
\def\suml{\sum_{\ell=0}^{\infty}}

\def\lfactor{{2\ell+1\over 4\pi}}
\def\lfactors{{(2\ell+1)/ (4\pi)}}

\def\lmin{\ell_{\min}}
\def\lmax{\ell_{\max}}
\def\kmin{k_{\min}}
\def\kmax{k_{\max}}

\def\lpix{\ell_{\rm pix}}

\title{On the linear filters for point source extraction of the {\it
Planck} mission}

\author[Lung-Yih Chiang \& Pavel Naselsky]
{Lung-Yih Chiang$^{1}$ and Pavel Naselsky$^{1,2}$\\
$^{1}$Theoretical Astrophysics Center, Juliane Maries Vej 30,
DK-2100,  Copenhagen, Denmark \\
$^{2}$Niels Bohr Institute, Blegdamsvej 17, DK-2100 Copenhagen,
Denmark}
\date{Accepted 2003 ???? ???; Received 2003 ???? ???}

\begin{document}
\maketitle
\newcommand{\etal}{{et~al.~}}
\begin{abstract}
The manifestation of point sources in the upcoming \planck maps is a direct
reflection of the properties of the pixelized antenna beam shape for each
frequency, which is related to the scan strategy, pointing accuracy, noise
properties and map-making algorithm. In this paper we firstly compare
analytically two filters for the \planck point source extraction,
namely, the adaptive top-hat filter (ATHF), the theoretically-optimal
filter (TOF). Our analyses are based on the premise that the experiment
parameters the TOF assume and require are already known: the
CMB and noise power spectrum and a {\it circular} Gaussian beam shape and size,
whereas the ATHF does not need any {\it a priori} knowledge. The
analyses show that the TOF is optimal in terms of the gain after the
parameter inputs. We simulate the {\it Planck} HFI 100 GHz channel
with elliptical beam in rotation to test the efficiency of the TOF and
the ATHF. We also apply the ATHF on the \wmap Q-band map and the
derived map (the foreground-cleaned map by Tegmark, de Oliveira Costa
\& Hamilton) from the \wmap 1-year data. The uncertainties on the angular power spectrum will hamper the efficiency of the TOF. To tackle the
real situations for the \planck point source extraction, most importantly, the
elliptical beam shape with slow precession and change of the ellipticity ratio
due to possible mirror degradation effect, the ATHF is
computationally efficient and well suited for the construction of
the \planck Early Release Compact Source Catalogue.

\end{abstract}

\section{Introduction}

The observation of cosmic microwave background (CMB) radiation by ESA
\planck mission will be able to provide the understanding of our 
Universe with many key cosmological parameters with unprecedented
precision. The signal received from the sky, however, will contain not
only the CMB radiation, but also foreground emissions such as
synchrotron, dust, free-free emission, Sunyaev-Zel'dovich effect from
galaxy clusters and extra-galactic point sources. Removing foreground
contaminations is therefore one of the main tasks for the \planck mission. Of
these different foreground emissions, the purpose for extraction of
extra-galactic point sources has two folds: to separate the
foregrounds from the underlying CMB signals to achieve the scientific
goal of the \planck mission as this type of foreground contamination
is rather non-Gaussian. On the other hand, the whole sky coverage and
a wide range of the \planck observing frequencies provide an
opportunity for producing the extra-galactic point source catalogue,
most notably, the proposed Early Release Compact Source Catalogue
(ERCSC), which will be released in roughly 7 months when the full sky
is covered once.

Recently, the NASA {\it Wilkinson
Microwave Anisotropy Probe} (\wmap) experiment \cite{wmap} has demonstrated
the importance of the component separation tools for extraction of CMB
power spectrum from the raw data. One of the sources with localized
peculiarities of the CMB sky is from the  point source
contamination. For the \wmap mission, the catalogue of the radio point
sources was obtained \cite{wmapfg,pierpaoli} and the final CMB maps
were cleaned from the point source contribution.

The \wmap channels at high frequency correspond to the \planck LFI frequency
range, in which the contamination of point sources seems 3 times higher
than that from the \wmap. For the \planck HFI channels, the point sources
are expected to be more significant regarding the issue of component
separation and of point source catalogue.

In this paper, we compare the theoretically-optimal
filter \cite{td98} (hereafter TOF) and the adaptive top-hat filter
\cite{tophat} (ATHF) proposed for extra-galactic point source
detection. Theoretically speaking, the filter proposed by
Tegmark \& de Oliveira-Costa \shortcite{td98} is {\it optimal} in
terms of the gain factor. The gain factor is an indicator of the amplitude
enhancement of point sources relative to the background. By
`theoretically' we mean that the required information input for this
filter is known: the CMB and the noise power spectrum, and the
circular beam shape and size.  This filter is known as a `Matched Filter' in
the field of signal processing.

A lot of effort has been devoted to the development of other linear
filters \cite{cayon,vielva1,pseudo,vio,barreiro,vielva2}, namely the
spherical Mexican Hat Wavelet (MHW), and the so-called
pseudo-filter. Vio et al. (2002) and  Barreiro et al. (2002) have
performed the comparison between the TOF and the pseudo-filter. Vielva et
al. (2001, 2003) have applied the MHW on simulation data for the
\planck mission.

The \planck in-flight beam shape properties are crucial in connection
with the  point source extraction. There are many issues regarding the
in-flight antenna beam shape properties and its reconstruction
\cite{burigana2001,fosalba,beam,degradation}. First of all, according
to optics calculations the main beams are not circular Gaussian
\cite{burigana2001}, but elliptical, which is normally formulated as
bivariate Gaussian functions. Secondly, with the most likely scan strategy, the
inclination angle of the beam is not all parallel through the sky due
to possible precession of its spin axis, but is rather slowly rotating
along circular scans. Note that this slow rotation of the antenna beam
is different from the balloon-borne experiments such as {\it
MAXIMA-1}, which has a fast rotating beam \cite{wu}, or {\it WMAP}, in
which the pixels have hits of multiple orientations \cite{wmapbeam},
and therefore a resultant quasi-symmetric pixel-beam. Furthermore,
there could exist mirror degradation effect during the approximately
15-month routine operation of the \planck mission \cite{degradation},
which will mimic the change of the inclination angle and the shape of
the beam. The combined effect from the above-mentioned factors will cause
the in-flight beam shape to change with time not only its ellipticity
ratio $\sigma_{+}/\sigma_{-}$, where $\sigma_{+}$ and $\sigma_{-}$
are the major and minor axis of the ellipse, but also
the inclination angle. There are also some minor effects such
as the issues on pointing accuracy and the noise properties that could also
mimic the beam degradation. The filters which assume a circular Gaussian beam
for point source extraction will need corresponding modifications. 

The adaptive top-hat filter is flexible for real situations. The ATHF
is similar to the so-called Gabor transform of the signal, using a
adaptive width of the kernel. The ATHF has a top-hat shape with two
cut-off scales working in harmonic domain. The two cut-off scales
$\lmax$ and $\lmin$ serve to cut down the unwanted power from both
sides of the power spectrum. Simultaneously, the two scales retain the
essential part of power spectrum where the convolution (by  the beam)
of point sources is most pronounced.

We also propose a median filtering technique for removing the point
source from the maps. The median filter can be used on combination
with any methods of point source detection in order to increase the
precision of the CMB map reconstruction.

The layout of this paper is as follows. In Section 2 we discuss the
antenna beam shape properties by firstly introducing the
\planck scan strategy and the definition of the window function. In
Section 3 we present the detailed analysis on the
theoretically-optimal filter (TOF) (the Matched Filter) and the adaptive
top-hat filter (ATHF). In Section 4 We compare by
simulations the TOF and the ATHF on gain capability on more realistic
situations and apply the ATHF on  Discussions and conclusion are in Section 5.

\section{The Planck antenna beam shape properties}
\subsection{The scan strategy}
The manifestation of point sources in the upcoming \planck maps is a direct
reflection of the properties of the pixelized antenna beam shape, which is
related to the scan strategy, pointing accuracy, map-making algorithm and the
extraction of the systematic effects from the time-ordered data (TOD)
set and pixelized maps as well. Before our analysis of point source
problem, in particular, for the \planck Early Release Compact Source
Catalogue (ERCSC), we need to describe the \planck
experiment when the TOD contain the information about the signal (and
noise) from a large number of circular time-ordered scans
\cite{dpa,beam}. Below we assume for simplicity that the systematic
features are already removed and the instrumental noise is Gaussian
white noise. In the temporal domain the observed signal $m_t$ is the
combined signal $d_t$ of the CMB and foregrounds from the sky
(hereafter beam-convolved `sky signals'), plus random instrumental noise $n_t$,
\begin{equation}
m_t=d_t + n_t
\label{eq:eq1}
\end{equation}
where 
\begin{equation}
d_t = \suml \summ B_{t,\ell m} \alm \Ylm ({\bi r}_t),
\label{eq:eq2}
\end{equation}
and $B_{t,\ell m}$ is the multipole expansion of the time-stream beam 
$B_t({\bi r}_t)$. In Eq.~(\ref{eq:eq2}) $\alm$ are the corresponding
multipole coefficients of the CMB and foreground signal expansion
on the sphere and $\Ylm$ are the spherical harmonics. 

Following Tegmark \& Efstathiou \shortcite{tegefs} we also assume that
map making algorithm is linear. Thus the signal in each pixel $s_p$
should have the following relation 
\begin{equation}
d_t= \sum\limits_{p=0}^{N}{\sf M}_{t,p} s_p,
\label{eq:eq3}
\end{equation} 
where ${\sf M}_{t,p}$ is the corresponding pointing matrix and $s_p$
is the sky signals convolved by the pixel beam $B_{p,\ell m}$, 
\begin{equation}
s_p=\suml \summ B_{p,\ell m} \alm \Ylm ({\bi x}_p),
\label{eq:eq4}
\end{equation}
${\bi x}_p$ is the two-dimensional vector with its components $x_p$
and $y_p$ denoting the location on the surface of the sphere. 
The definition of the pointing matrix ${\sf M}_{t,p}$ depends on the
scan strategy of an experiment. Below, as a basic model, we will use the
model of the scan strategy of the \planck mission which was discussed
by Burigana et al. \shortcite{burigana1998}. Let ${\bi s}$ be the unit vector
along the satellite spin-axis in the anti-Sun direction and ${\bi o}$ is
that along the direction of the optical axis of the telescope. The
angle between spin-axis and optical axis is $\alpha_0=\cos^{-1}({\bi
s}\cdot {\bi o}) \simeq 85^{\circ}$. The satellite itself will scan
the same circle 60 times around the spin-axis at $\Omega=1$
r.p.m.. Each hour the spin-axis is manoeuvred along the ecliptic plane
by $\sim2\farcm5$. Note that for our geometrical model, during the
circling of the optical axis around the spin-axis in each step, the
inclination of the beam relative to the optical axis are stable.  For
modelling of the \planck antenna beam shape we will describe the simplest scan
strategy with the spin axis right on the ecliptic plane without any
additional (regular) modulation (e.g. it can be above or below the
plane due to precession). This model reflects
the geometrical properties of the beam asymmetry and their
manifestation in the pixelized maps. Under the assumptions mentioned
above we can now describe the elliptical beam shape model as follows, 
\begin{equation}
B_t({\bi x}-{\bi x}_t)= 
\exp \left[-\frac{1}{2} ({\sf RU})^{\rm T} {\sf D}^{-1} ({\sf RU})\right],
\label{eq:eq5}
\end{equation}
where 

\begin{equation}
{\sf U}=\left(
\begin{array}{c}
        x-x_t \\
        y-y_t

\end{array}
\right),
\end{equation}
${\sf R}$ is the rotation matrix which describes the inclination of
the elliptical beam,
\begin{equation}
{\sf R}=\left(
\begin{array}{rr}
       \cos \alpha  &  \sin \alpha  \\
      -\sin \alpha  &  \cos \alpha 
\end{array}
\right), \label{eq:inclination}
\end{equation}
with $\alpha$ being the inclination angle between $x$ axis and the
major axis of the ellipse. The ${\sf D}$ matrix denotes the beam
dispersion along the ellipse principal axes, which can be expressed as
\begin{equation}
{\sf D}=\left(
\begin{array}{cc}
      \sigma^2_{+} &     0          \\
          0        &  \sigma^2_{-}

\end{array}
\right). \label{eq:dispersion}
\end{equation}
The center of the Cartesian coordinate system is denoted by $x_t$ and
$y_t$ at some moment $t$ with $z$ axis along
the ${\bi o}$, and $x$ and $y$ axis on the plane tangent to the celestial
sphere. We will choose below the standard orientation of the local
Cartesian coordinate system with $x$ axis parallel to the scanning
direction.

After some time $\Delta t=2\pi N_{\rm rot}/\Omega$, where
$N_{\rm rot}=60$ is the number of the sub-scans, the spin axis
of the satellite shall be stepped along the ecliptic plane to ${\bi s}^{'}$ with
the small deviation $\Delta{\bi s} = {\bi s}^{'} - {\bi s} \sim 2\farcm5$. 
For a small part of the sky we can neglect
the rotation of the Cartesian coordinate system and choose the $x^{'}$
and $y^{'}$ axes parallel to $x$, $y$ axis.

The received signals in
each pixel $p$ depend on the orientation of the pixel beam
$B_p({\bi x})$ and the location of its center \cite{wu}. 
For the upcoming \planck non-symmetric spatially-dependent beam, the
convolution of the signal with asymmetric beam immediately produces
non-Gaussian signature coupled with the underlying signal, which affects the
estimation of angular power spectrum. Although the real beam profile 
(including the sidelobes) are complicated, we begin our analysis
from the model of the time-dependent, elliptical beam shape which
is applicable up to $\sim \,-30$ dB level. At low amplitude part of the
beam shape ($<-30$ dB) the complexity of the beam estimation increase
dramatically. Moreover, due to possible precession of the spin axis
${\bi s}$ (around the standard orientation), which can be described as some
additional noise, it is possible to find some transformation of the
width of the beam taking into account the statistical properties of the
precession. In such a case the angle between the spin axis and the optical
axis should be $\phi(t)=\phi_0 + \nu(t)$, where $\phi_0\sim
85^{\circ}=\const$ and $\nu(t) \ll \phi_0$ denote a random precession. 

\subsection{The window function}
Due to the scan strategy and (possible) precession, the pixel
beam pattern, in principle, should be different for different pixels
on the sphere, which means that we need some modification of the standard
tools for the $\Cl$ power spectrum extraction by taking into account
any possible peculiarities from the bivariate Gaussian beam shape. The finite
resolution of the pixel beam means that the signal we receive from the sky is
smoothed by the beam,
\begin{equation}
T({\bi p})=\int d\Omega_{\bi n} B({\bi p},{\bi n})T_{\rm CMB}({\bi n})
\end{equation}
Usually the influence of the pixel beam on the observed signals can be
described in terms of the window function \cite{white,souradeep}
\begin{equation}
W_{\ell,({\bi p},{\bi q})}^2=
\frac{4\pi}{2\ell+1} \summ B_{\ell m,{\bi p}} B^{*}_{\ell m,{\bi
q}} 
\label{eq:eq6}
\end{equation}
where ${\bi p}$ and ${\bi q}$ denote the positions of the corresponding
pixels in the map and 
\begin{equation}
B_{\ell m,{\bi p}}= \int d \Omega_{\bi n} B({\bi p}, {\bi n})
\Ylm^{*}({\bi n})
\label{eq:eq8}
\end{equation}
is the spherical harmonic transform of the beam function pointing in
the direction ${\bi p}$. This window function determines the CMB signal correlation matrix as 
\begin{equation}  
C_{({\bi p},{\bi q})}=\suml \lfactor \Cl
W_{\ell,({\bi p},{\bi q})}^2
\label{eq:eq7}
\end{equation}
In the framework of the \planck mission it is possible to use the flat
sky approximation (as a part of the whole sky $\Cl$ harmonic analysis)
in order to estimate the point source contamination and beam influence
on the high multipole part of the $\Cl$ power spectrum. For such flat
sky limit the computation cost is minimal, as we can implement Fast Fourier
Transforms (FFT) for the CMB anisotropy. When the sky signal is
convolved with an elliptical beam with an angle between its major axis and the
scan direction, the temperature becomes 
\begin{equation}
T({\bi p}) = \int_0^{\infty}
\frac{d{\bi k}}{(2\pi)^2} e^{i {\bi k} \cdot {\bi p}} B({\sf R}_{\alpha}[{\bi k}]) T_{\rm
CMB}({\bi k}), 
\label{eq:eq9}
\end{equation}
where $B$ is the Fourier transform of the beam profile at the
origin, and ${\sf R}_{\alpha}[{\bi k}]$ is the rotation operator with an angle $\alpha$
with respect to the scan direction \cite{souradeep}, which can be written
explicitly as
\begin{equation}
B({\sf R}_{\alpha}[{\bi k}])=\exp\left[-\frac{1}{2}({\sf R K})^{\rm T} {\sf D}({\sf R K}) \right
],
\end{equation} 
where ${\sf R}$ and ${\sf D}$ are defined as in Eq.~(\ref{eq:inclination}) and
(\ref{eq:dispersion}) and 
\begin{equation}
{\sf K}=\left(
\begin{array}{c}
        k_x \\
        k_y
\end{array}
\right).
\end{equation}
The window function now can be expressed as \cite{souradeep}
\begin{eqnarray}
\lefteqn{W^2_{\bi k}(\alpha, \Delta \theta) =}\nonumber \\
&&\int_0^{2\pi}\frac{d\phi_k}{2\pi}
e^{i {\bi k} \cdot \Delta {\bi \theta}}
B({\sf R}_{\alpha}[{\bi k}]) B^{*}({\sf R}_{\alpha+\Delta \theta}[{\bi k}]),
\end{eqnarray}
where $\phi_k$ is the angle of wave vector ${\bi k}\equiv(k_x,k_y)$ in the ${\bi k}$ space such
that $k_x= k \cos \phi_k $ and $k_y=k\sin \phi_k $.
Moreover, for the flat sky approximation (and without the influence of the
instrumental noise and any peculiarities of the scanning such as
precession) we can use the assumption of stable beam orientation,
$\alpha=\const$, with the corresponding Fourier image for the elliptical
part of the beam shape,
\begin{eqnarray}
B({\bi k}) & = & \exp \left[-\frac{\sigma^2_{+}(k_x \cos \alpha + k_y \sin
\alpha)^2}{2} \right. \nonumber \\
&-& \left. \frac{\sigma^2_{-}(-k_x \sin \alpha + k_y \cos
\alpha)^2}{2}\right] \nonumber \\
& = & \exp \left[-\frac{k^2_x}{2a^2}-
\frac{(k_y-k_x\frac{b^2}{a^2})^2 a^2\sigma^4}{2}\right]
\label{eq:eq10}
\end{eqnarray}
where $\sigma^2=\sigma_{+}\sigma_{-}$, $r=\sigma_{+}/\sigma_{-}$,
$a^2=(\cos^2 \alpha +r^2 \sin^2 \alpha)/r\sigma^2$ and
$b^2=-(r^2-1)\sin 2\alpha / 2r\sigma^2$. 
Without loss of generality, we can rotate the coordinate system such that
$\alpha=0$, and, for small $\Delta \theta$, $\Delta {\bi \theta}\equiv {\bi
\theta}=(\theta_x, \theta_y)$ with $\beta = \arctan(\theta_y / \theta_x)$. The
window function can be written as 
\begin{eqnarray}
W_{\bi k}^2 (\Delta \theta) \equiv &  W_{k}^2(\theta, \beta) =
\int_0^\pi \frac{d\phi_k}{\pi} \cos [ k \theta \cos(\phi_k-\beta) ] \nonumber \\  
\times &\exp \left\{ -k^2\sigma^2 r
\left[ 1+ (r^{-2}-1) \sin^2 \phi_k \right] \right\},
\label{eq:eq12}
\end{eqnarray}
with the bivariate beam profile 
\begin{equation}
B({\bi k})= \exp\left\{-\frac{k^2\sigma^2 r}{2}
\left[1+(r^{-2}-1)\sin^2 \phi_k \right] \right\}.
\label{eq:eq11}
\end{equation}
The integral in Eq.~(\ref{eq:eq12}) can be expressed analytically
\cite{souradeep}. If the asymmetry of the beam shape is low then we can express
$W_{k}(\theta)$ as an infinite series in powers of asymmetry parameter
$\epsilon_k = k^2 \sigma^2(r^{-2}-1)$, 
\begin{equation}
W_{k}^2(\theta, \beta)=\sum\limits_{n=0}^{\infty}(-1)^n(\epsilon_k)^nw_{k,n}
\end{equation}
where 
\begin{eqnarray}
w_{k,n} = B \left(\frac{1}{2},n+\frac{1}{2}\right) (\cos\beta)^{2n}
\exp(-k^2\sigma^2 r ) \nonumber \\
\times \sum \limits_{m=0}^{n}
\frac{(\tan \beta)^{2m} F_2\left( \frac{1}{2} + m,\frac{1}{2}, 1+n;
-k^2 \theta^2/4  \right)}{\Gamma(n-m+1)\Gamma(m+1)},
\label{eq:eq13}
\end{eqnarray}
$\Gamma(a)$ and $B(a,b)$ are the Euler Gamma function of the
first and the second kind, respectively, and $F_2$ is a generalized
hyper-geometric function. For beams with symmetric shape where $r=1$ and
$\sigma_{+}=\sigma_{-}=\sigma$, we obtain from Eq.~(\ref{eq:eq13})
\begin{equation}
W_{k}^2(\theta)=J_0(k\theta)e^{-k^2\sigma^2},
\end{equation}
where the zeroth-order Bessel function $J_0(k\theta)$ corresponds to
the standard asymptotic of the Legendre polynomials
$P_{\ell}(\cos \theta)\approx J_0(k\theta)$ at $k=\ell+1/2$. 
In a general case when $r\ge 1$ the window function $W_k^2(\theta)$ is a function of the separation angle $\theta$ and phase $\beta$ as
from Eq.~(\ref{eq:eq13}). This fact reflects the influence of the anisotropy 
of the beam on the CMB signal after convolution which transforms isotropic
CMB signal to anisotropic one.

For applications of the above-mentioned properties of the beam shape
and the window function for point source extraction, let us define the mean
beam profile as follows,
\begin{equation}
\overline{B}_{\ell}=\frac{1}{2\ell+1} \summ \Blm. \label{eq:B}
\end{equation}
In the flat sky approximation, taking Eq.~(\ref{eq:eq11}) into
account, the mean beam profile is related to the ellipticity parameter $r$ as 
\begin{equation}
\overline{B}(k) = \exp\left[-\frac{k^2\sigma^2(r^2+1)}{4r}\right]
I_0\left[ \frac{k^2\sigma^2(r^2-1)}{4r}\right], 
\label{eq:eqmb}
\end{equation}
where $I_0(x)$ is the Bessel function of the first kind.
For the window function $W^2(k)$ in the flat sky approximation we have 
\begin{eqnarray}
W^2(k) = \int_0^{\pi} \frac{d\phi}{\pi} \exp\left\{-\frac{k^2\sigma^2}{r}
\left[1+(r^2 -1) \sin^2 \phi \right]\right \} \nonumber \\
=\exp\left[-\frac{k^2\sigma^2(r^2+1)}{2r} \right]
I_0\left[ \frac{k^2\sigma^2(r^2-1)}{2r}\right]. 
\label{eq:eqm1}
\end{eqnarray}
The last two approximations play a crucial role in point source extraction.
First of all, the actual value of the beam profile at the location of
the point source is related to the orientation of the beam as 
\begin{eqnarray}
B({\bi k})& = & \overline{B}(k) I_0^{-1}
\left[ \frac{k^2\sigma^2(r^2-1)}{4r} \right] \nonumber \\
& \times & \exp\left[-\frac{k^2\sigma^2 (r^2-1)}{4r}
\cos 2\phi_{{\bi k}}\right],
\label{eq:eqm2}
\end{eqnarray}
and is different from the mean beam profile. These differences depend on
the position of the point source in the map and for a given value of
the flux $S_i$ the result of the point source subtraction should be
different for different points of location.

\section{Optimization of the linear filters} \label{calculations}
In this section we would like to elaborate for point source extraction the
subtleties of the 3 linear filters: the theoretically-optimal filter, the
adaptive top-hat filter and the pseudo-filter. The analyses and comparisons are
based on the assumption that the beam shapes for all frequency channels are
circular Gaussian and are known if
required (by the TOF and the pseudo-filter) and the signal properties, both the
CMB and pixel noise power spectrum, are known if required (by
the TOF). The analysis and comparison is in terms of the gain factor. The gain factor is
defined as $R_{\rm f}/R_{\rm i}$, where $R_{\rm f}$ is the signal ${\rm S}_{\rm
f}$ to noise ${\rm N}_{\rm f}$ ratio with ``${\rm f}$'' indicating after convolution by the
filter.  

\subsection{The theoretically-optimal filter}
In the following analyses for the theoretically-optimal filter (TOF) we adopt
the convention of notation by Tegmark \& Efstathiou \shortcite{tegefs}, Tegmark
\& de Oliveira-Costa \shortcite{td98} (hereafter TD98), i.e., the
{\it de-convolved} power spectrum of the combined signal is defined as  
\begin{equation}
\Cl=\Cl^{\rm sky}+\frac{C^{\rm pix}}{W_\ell^2},
\label{eq:ctot}
\end{equation}
where $\Cl^{\rm sky}$ is the sky signals including CMB and all
foregrounds but the point source contribution, $C^{\rm pix}$ is the noise power spectrum, which
is taken as $({\rm FWHM} \sigma_{\rm pix})^2$, and $W_\ell^2$ is the window
function. Thus the {\it combined} (except point sources) power spectrum in
the map is $\Cl^{\rm tot-ps} \equiv \Cl W_{\ell}^2$. Note that the in-flight
beam $\Blm$ and the window function $W_\ell^2$ definitely
have some error bars caused by possible mirror degradation effects, the
galactic foreground contaminations and the instrumental noise. Thus simple
inversion of $W_\ell^{-2}$ for the power spectrum requires specific
renormalization due to the antenna beam shape properties.

The idea to construct the theoretically-optimal filter (TOF)
 for point source extraction is outlined in TD98. We can write down the 
signal as the sum of the point source contribution and the rest (TD98)
\begin{equation}
x(\rh)= g \sum_i S_i  \delta_D(\rh_i, \rh)+ \sumlm \alm \Ylm(\rh),
\end{equation}
where $\delta_D$ is a Dirac delta function, $S_i$ is the flux of the point
source at the direction $\rh_i$, $\alm$ is the coefficients of the signal
decomposition with $\langle |\alm|^2 \rangle \equiv \Cl$ as in
Eq.~(\ref{eq:ctot}), $\Ylm(\rh)$ the spherical harmonics, and
\begin{equation} 
g=\frac{1}{2k} \left(\frac{hc}{k T_{\rm CMB}}\right)^2 \frac{[2 \sinh
(h \nu/2 k T_{\rm CMB})]^2} {(h\nu/k T_{\rm CMB})^4}. 
\end{equation}
This filter is named as `theoretically-optimal' because it is constructed in
such a way that the gain factor will be maximal, i.e., when the `observed'
signal, $ B \otimes x(\rh)$ is convolved with the filter $F$,
\begin{eqnarray} 
y(\rh) & = & g \sum_i S_i(F \otimes B)(\rh_i \cdot \rh)\nonumber \\
& + & \sumlm \Flm \Blm \alm \Ylm(\rh),
\label{eq:filterconvolved}
\end{eqnarray} 
the maximization of the `signal' to `noise' ratio
(${\rm S}_{\rm f}/{\rm N}_{\rm f}$) will give us the filter shape, where ${\rm S
}_{\rm f}$ is the square root of the variance of the
point source amplitudes and ${\rm N}_{\rm f}$ now is the {\it rms} of the
combined signal, both convolved by the filter $F$. \footnote{Note that
this definition of our signal-to-noise ratio is inverse of that in TD98.} Equivalently, we
want to maximize  
\begin{eqnarray}
R & =  & \frac{g^2 \langle \sum_i \sum_j S_i S_j [(F \otimes B)(\rh_i
\cdot \rh)] [(F \otimes B) (\rh_j \cdot \rh)] \rangle }{\sum_\ell \lfactors
\Fl^2 \Wl^2 \Cl} \nonumber \\   
& = &  g^2\sum_i S_i^2 \frac{\left[\sum_\ell \lfactors \Fl
\overline{\Bl}  \right]^2}{\sum_\ell \lfactors \Fl^2 \Wl^2 \Cl}, 
\label{eq:G}
\end{eqnarray}
where $\otimes$ denotes convolution, $\Blm$ are the coefficients of the
spherical harmonic decomposition of the beam profile, $\Flm$ is the multipole
expansion of $F$, and $\overline{\Bl}$ is the mean beam. We thus can obtain
the TOF for each \planck frequency channel. 


 
Note that in reaching Eq.~(\ref{eq:G}) the filter is assumed isotropic in the
Fourier rings and statistically homogeneous. Thus in the spherical harmonic
decomposition this filter has only $\ell$-dependence, but not on the
azimuthal numbers $m$. This condition is only valid when the sky signal and
the pixel noise are Gaussian. If the sky signal (or the pixel noise) is non-Gaussian, however,
then Eq.~(\ref{eq:G}) needs the corresponding modification. For the following
analyses we assume for simplicity that the Gaussian assumption is appropriate
 for the CMB plus foreground. 

The determination of the TOF is similar to the definition of
functional derivatives definition. Suppose that $\overline{\Fl}$
corresponds to the maximization of the ratio $R$ from
Eq.~(\ref{eq:G}), then a small variation $\fl$ of the shape from
the $\overline{\Fl}$ shall shift the function
$R(\overline{\Fl} +\fl)$ away from the maximal value $R_{\max}$, which
is proportional to the $\fl$, $\fl^2$\ldots, if $||f||\ll 1$.  So we
have 
\begin{equation}
\Fl=\overline{\Fl}+ \fl,
\label{eq:f}
\end{equation}
and obtain the following formulae from Eq.~(\ref{eq:G}),
\begin{equation}
\Delta \equiv R(\Fl)-R(\overline{\Fl})= g^2\sum_i S_i^2
\frac{M+Q}{\alpha P},
\label{eq:D}
\end{equation}
where
\begin{equation}
M= 2\beta \sum_n \frac{2n+1}{4\pi} \left[\alpha \overline{B}_n -
\beta W^2_n C_n \overline {F}_n\right] f_n,
\label{eq:M}
\end{equation} 
\begin{equation}
\alpha=\sum_\ell \lfactor \Wl^2\overline{\Fl}^2 \Cl,
\label{eq:a}
\end{equation}
\begin{equation}
\beta=\sum_\ell \lfactor \overline{\Bl} \,\overline{\Fl},
\label{eq:b}
\end{equation}
\begin{eqnarray}
Q & = &  \alpha \left(\sum_n \frac{2n+1}{4\pi} \overline {B}_n f_n\right)^2
\nonumber \\
& - &\beta^2 \sum_n \frac{2n+1}{4\pi} W^2_n f^2_n C_n, 
\label{eq:N}
\end{eqnarray}
and
\begin{equation}
P = \sum_\ell \lfactor \Wl^2 (\overline{\Fl}+\fl)^2 \Cl.
\label{eq:P}
\end{equation}

Equivalently, to maximize the gain factor $R(\Fl)$ is to take $M=0$,
the condition that the first functional derivative is zero,
$\delta R/\delta \Fl=0$. If we assume that the TOF profile
is an analytic function, then from Eq.~(\ref{eq:M}) we get
\begin{equation}
\overline{F}_n=\frac{\alpha\overline{B}_n}{\beta W^2_n C_n}.
\label{eq:F}
\end{equation}
As one can see from Eq.~(\ref{eq:a}) and (\ref{eq:b}) the
coefficients $\alpha$ and $\beta$ are also the functionals of the TOF
profile. Substituting Eq.~(\ref{eq:F}) into
Eq.~(\ref{eq:a}) and (\ref{eq:b}) we obtain the following
relationship between $\alpha$ and $\beta$ functionals, 
\begin{equation}
\beta^2=\alpha \sum_\ell \lfactor \frac{\overline
{\Bl}^2}{\Wl^2 \Cl}
\label{eq:ab}
\end{equation}
Taking into account the definitions of the $\alpha$ and $\beta$
functionals and Eq.~(\ref{eq:ab}), we reach the conclusion that
the TOF $\overline{F}_n$ from Eq.~(\ref{eq:F}) have the form
\begin{equation}
\overline{\Fl}_=\frac{\rho\overline {\Bl}}{\Wl^2 \Cl},
\label{eq:ro}
\end{equation}
where $\rho=\alpha/\beta$ and $(\beta^2/\alpha) g^2\sum_i
S_i^2 $ is the maximal gain after filtering,
\begin{equation}
R^{\rm TOF} \equiv R(\overline{\Fl}) = g^2\sum_i S_i^2\sum_\ell
\lfactor \frac{\overline{\Bl}^2}{\Wl^2 \Cl}.
\label{eq:go}
\end{equation}
If the TOF $\overline{\Fl}$ corresponds to the global
maximum of the gain (in the class of the analytic functions) then
the second functional derivative $\delta^2 R/\delta \Fl^2$ should
be negative at  $\Fl=\overline{\Fl}$. For small
variations around $\overline{\Fl}$ this corresponds to the
condition $Q<0$, according to Eq.~(\ref{eq:D}). After substituting
$\fl=\overline{\Fl} \; \epsilon_\ell$ into Eq.~(\ref{eq:N}) we get
\begin{eqnarray}
Q & = & \alpha \left(\sum_\ell\sum_n h_\ell h_n \epsilon_\ell \epsilon_n -
\sum_n h_n \sum_\ell h_\ell \epsilon^2_\ell \right)\nonumber \\
& = & -\frac{1}{2} \sum_\ell \sum_n h_\ell h_n ( \epsilon_\ell - \epsilon_n)^2,
\label{eq:delta}
\end{eqnarray}
where $h_\ell=\lfactors \overline{\Bl}^2/ \Wl^2 \Cl$.
As one can see from Eq.~(\ref{eq:delta}), for any values of the
$\epsilon_\ell$ functions, the function $Q$ is negative and the filter
$\Fl=\overline{\Fl}$ corresponds to the maximal gain in the functional space of the linear filters. 

The filter shape of Eq.~(\ref{eq:ro}) for point source extraction is a
generalization of the TD98 filter, which is obtained under the assumption of 
circular Gaussian antenna beam shapes. The new element which we get
for the more complicated \planck antenna beam shape in
Eq.~(\ref{eq:ro}) shows that for the TOF construction,
instead of the TD98 model, we need to know the mean beam (over all $m$ from
Eq.~(\ref{eq:B}) beam profile), the window function $\Wl^2$, and all the
signal properties as well. 

One of the most crucial factors affecting the efficiency of the TOF is 
from the systematic effects, namely, the possible degradation
effect of the primary and secondary mirror surface during the flight
\cite{degradation}. As mentioned in the Introduction, the mirror degradation 
could change the ellipticity ratio and the beam size, so the calibration and
reconstruction of the \planck in-flight antenna beam shape will have direct
consequence of Eq.~(\ref{eq:B}) and the window function.

From a practical point of view, when we
apply the TOF of Eq.~(\ref{eq:ro}) for the \planck point source extraction,
in particular for the ERCSC, it is necessary to know precisely the power
spectrum of the sky signals and the pixel noise properties. It is, however,
only possible at the earliest after the analyses
of component separation and $\Cl$ extraction when the mission is completed. 
Conversely, for the realization of the $\Cl$ extraction it is
necessary to know the window function $\Wl^2$ for all frequency
ranges.\footnote{One plausible idea is to use very bright point sources with
flux above 1 Jy for determination of the window function and the
mean beam shape at any time of the observations (see, e.g., Chiang \etal
2002a).} This means that for the ERCSC construction it is better to 
to use primitive filters which require as little information as possible
about the properties of the antenna beam shape and the sky signals.

One additional possibility for the new class of the linear filters
comes from Eq.~(\ref{eq:M}) at $M=0$ but for the non-analytical shape
of the optimal filter $\overline {\Fl}$, which is not in the scope of this
paper. In the next section we will demonstrate that the maximization
of the gain factor can be obtained using a simple top-hat filter.

\subsection{The Adaptive Top-Hat filter (ATHF)}
Chiang et al. \shortcite{tophat} introduce another class of filter similar to the Gabor transform \cite{gabor,gabortrans}: the adaptive top-hat filter 
(ATHF), which is implemented in the harmonic space with two adaptive cut-off
parameters, $\Fl^{\rm TH} = \Theta(\ell-\lmin)\Theta(\lmax-\ell)$,
where $\Theta$ is a Heaviside function, $\lmin$ and $\lmax$ are two
adaptive parameters. The ATHF allows us to investigate a new class of
non-analytical functions for the ERCSC. Point
source extraction by the ATHF is based on the idea that each point
source in the  map (and the TOD as well) manifests itself as a
non-Gaussian feature which is convolved with the beam shape. The goal,
therefore, is to distinguish these non-Gaussian features from the CMB
signal and pixel noise by using some general properties of the CMB map
only. The ATHF is generalized from the amplitude-phase analysis method
for point source extraction in the \planck maps
\cite{amplitudephase}. In filtering, the parameter $\lmin$ suppresses
all the power of the signal from the sky at the multipole range $\ell
< \lmin$ for which the beam shape influence on the signal is not
crucial. This part of filtering by the ATHF reflects the fact that the
pure CMB signal has $\ell^{-2}$ tail, whereas dust, synchrotron and
free-free emission  have  $\sim \ell^{-3}$ tails for the foregrounds
\cite{tegefs}. The parameter $\lmax$, on the other hand, suppresses
the high multipole power from pixel noise and secondary anisotropies,
in order to relatively increase the point source amplitudes above the
threshold of detection. 

Non-symmetric beam break the statistical isotropy of the Fourier ring in the
power spectrum and the ellipticity of the
beam will manifest itself in the Fourier domain \cite{beam}. The
usual estimation of the ellipticity ratio of \planck antenna beam
($\sim 1.2\div 1.3$) is well above the ratio of the two
proposed cut-off scales of the ATHF for all the \planck channels
\cite{tophat}, which means the cut-off scales of the ATHF will retain
the part of the elliptic shape on the Fourier ring. 

Before our analyses, we would like to emphasize again the adaptiveness
of the ATHF which does not at all
need any {\it a priori} information of the experiment parameters. In principle,
we can start from any values of the
$\lmin$ and $\lmax$ parameters for investigation of the peak
amplitudes in the CMB map without any assumptions on the properties of
the signal and noise. We can choose, for example, $\lmax^{(0)}=\lpix$,
where $\lpix$ is the multipole mode corresponding to the pixel size, and
$\lmin^{(0)}=2$, the lowest multipole mode. After the first step of
the iteration, the maxima might contain both CMB peaks and point sources above
some specified threshold $\nu_t\sigma_f$, where $\sigma_f$ is the
square root of the variance of the filtered map and
$\nu_t$ is a dimensionless amplitude, which, following the standard
criterion (see, e.g. TD98), we will set $\nu_t=5$. 

The next few steps of the iteration are to increase the $\lmin$
parameter. Because the power of the low multipole part of the CMB and
foreground signal are gradually subtracted by the increasing
$\lmin$, the new variance of the signal after filtering
$\sigma^2_0$ must be smaller than the preceding one, but
the point source amplitudes changes only slightly, which results in
pushing the gain factor higher. The steps of increasing $\lmin$ are
carried on up to the step where the gain factor dips after
reaching a maximum value. We can then tune the $\lmax$ parameter in
the similar way by lowering it from $\lmax^{(0)}=\lpix$. 

Tuning of the cut-off scales of the ATHF will produce the
so-called Brownian motion of the CMB peaks, as compared to motions within the
beam size for beam-convolved point sources \cite{brownian}, which is useful in
lowering the threshold of the criterion for point source detection.  

For practical implementation of the iteration scheme, however, it is
not necessary to start from $(\lmin^{(0)}, \lmax^{(0)})= (2,
\lpix)$. In order to minimize the time needed for the realization of
the iteration scheme, we can use as the first step from the suggested
sets of $\lmin$ and $\lmax$ for all the \planck channels \cite{tophat}.
Below to obtain the maximal gain from the ATHF, we can give the same
treatment as that for the TD98 filter, but treating $\kmin$ and
$\kmax$ as variables. Again, the same assumptions are made on the
properties of the signal and noise similar to the TD98 filter and its
generalization TOF [Eq.~(\ref{eq:F})], namely, the mean beam, the
window function and the power spectra of the signal and noises are
known. For analysis we will use the flat sky approximation for a small
path of the sky. The generalization of the analysis
for the whole sphere can be done without special modification. 

Let us start with Eq.~(\ref{eq:G}) for the
flat path of the sky, the cut-off shape of the ATHF transforms the $\kmin$
and $\kmax$ parameters on to the summations at both numerator and
denominator,
\begin{equation}
R(\kmin,\kmax) =  g^2\sum_i S_i^2
\frac{\left[\int\limits_{\textstyle \kmin}^{\textstyle \kmax} dk\; k
\; \overline{B}(k) \right]^2}
{\int\limits_{\textstyle \kmin}^{\textstyle \kmax} dk\; k \;W^2(k) C(k)}, 
\label{eq:G1}
\end{equation}
where the shape of the filter $\overline{F}(k)= \Theta(k-\kmin)
\Theta(\kmax-k)$ now is quite peculiar. By the definition, the
perturbations of the $\kmin$ and $\kmax$ parameters, $\delta
\kmin$ and $\delta \kmax$, produce the perturbations of the filter
shape
\begin{eqnarray}
f(k) & = & \Theta(k - \kmax + \delta \kmax) \Theta(\kmax - k) \nonumber \\
& + &  \Theta(k- \kmin + \delta \kmin) \Theta( \kmin - k )
\label{eq:G2}
\end{eqnarray}

\begin{figure}
\centering
\epsfig{file=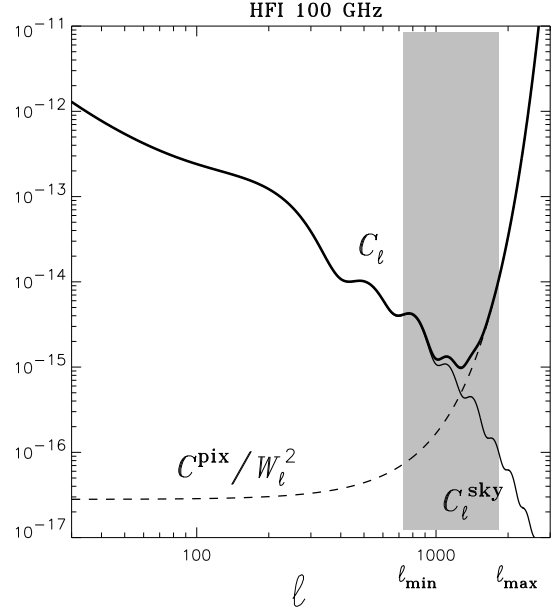,width=8cm}
\caption{The simulated angular power spectrum of the CMB and pixel noise for
the \planck HFI 100 GHz channel. The solid curve is the power spectrum
$\Cl^{\rm sky}$, which includes the CMB signal and foregrounds, and
the dash line is the {\it de-convolved} noise power spectrum $C^{\rm
pix}/\Wl^2$. The thick curve represents the {\it de-convolved} power
spectrum of the combined signal $\Cl$. The shaded area shows the
optimal top-hat filtering range $(\lmin, \lmax)$ for this channel,
taken from Chiang \etal (2002b). We can see from the figure that
$\Cl(\lmax) \simeq C^{\rm pix}/W^2(\lmax)$ and $\Cl(\lmin) \simeq C^{\rm
sky}(\lmin)$.} \label{powerspectrum} 
\end{figure}

As one can see from Eq.~(\ref{eq:G2}) these perturbations correspond to 
$||f(\ell)|| \gg ||F(\ell)||$ at the ranges $ \kmax -\delta \kmax \le k
\le \kmax$ and $\kmin - \delta \kmin \le  k \le \kmin$. Thus finding the
optimal values of the $\kmin$ and $\kmax$ parameters through the same treatment
as in the last section is not appropriate. From a theoretical point of view that
means that for a given top-hat shape of the filter the point of maxima of the
gain parameter does not correspond to any solutions for Eq.~(\ref{eq:M}). 

In order to find the optimal $\kmin$ and $\kmax$ values we use instead
the following conditions $\partial R( \kmin,\kmax)/ \partial \kmin=0$
and  $\partial R(\kmin,\kmax)/\partial \kmax = 0$. For 
Eq.~(\ref{eq:G1}) these conditions lead to 
\begin{eqnarray}
\lefteqn{2\overline{B}(\kmax)\int\limits_{\textstyle \kmin}^{\textstyle \kmax}dk\; k\; W^2(k)
C(k)}\nonumber \\
& = & W^2(\kmax)C(\kmax) \int\limits_{\textstyle \kmin}^{\textstyle
\kmax} dk\; k \; \overline{B}(k),
\label{eq:G3}
\end{eqnarray}
and 
\begin{eqnarray}
\lefteqn{ 2\overline{B}( \kmin )\int\limits_{\textstyle \kmin}^{\textstyle \kmax}dk\; k\; W^2(k)
C(k)}\nonumber \\
& = & W^2(\kmin) C(\kmin) \int\limits_{\textstyle \kmin}^{\textstyle
\kmax}dk\; k \; \overline{B}(k).
\label{eq:G4}
\end{eqnarray}
We get from Eq.~(\ref{eq:G3}) and Eq.~(\ref{eq:G4}) 
\begin{equation}
W^2(\kmin)C(\kmin)=\frac{\overline {B}(\kmin)}
{\overline {B}(\kmax)} W^2(\kmax) C(\kmax). 
\label{eq:equiv0}
\end{equation}
Apart from a trivial solution in Eq.~(\ref{eq:equiv0}), with undefined value of
the $\overline {k}=\kmax=\kmin$, there are non-trivial solutions. For 
comparison with other filters we use the same circular Gaussian model for the
antenna beam and the window function, $W^2(k)=\overline {B}^2(k)=\exp
(-k^2\theta_B^2)$, where $\theta_B$ is the ${\rm FWHM}/\sqrt{8 \ln
2}$, so from Eq.~(\ref{eq:equiv0}) we have $\overline{B}(\kmax)
C(\kmax)=\overline{B}(\kmin) C(\kmin)$. We can assume that $\kmin \theta _B <1$,
\begin{equation}
\overline{B}(\kmin)=\exp(-\kmin^2 \theta^2_B/2) \simeq 1 \gg \exp(-\kmax^2 \theta^2_B/2).
\label{eq:asum}
\end{equation}

\begin{figure}
\centering
\epsfig{file=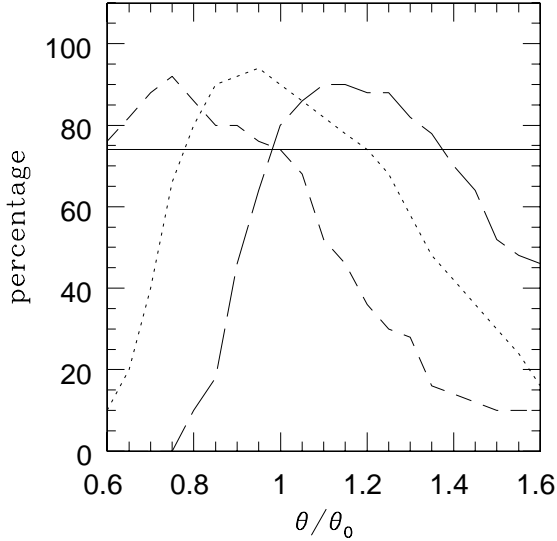,width=8cm}
\caption{The point source extraction rate from the TOF and the
ATHF. We put 50 point sources with $1.5 \sigma_{\rm CMB}$ randomly in the CMB map
simulating the \planck HFI 100 GHz channel with the corresponding pixel noise.
This number of point sources is too large for the 100 GHz frequency channel, it
nevertheless serves as the efficiency (percentage) of extraction rate. 
The elliptical beam is rotating across the simulation patch with ellipticity
ratio $\theta_{+}/\theta_{-}=1.3$.   
The plot is the extraction percentage of point sources against the $\theta$
inserted to the TOF (in terms of $\theta_0 = 10.7 /\sqrt{8 \ln 2}$ (arcmin) ).
In the dotted curve, the power 
spectrum is precisely known. The short-dash curve is the case when the slope of the CMB power
spectrum has 5 per cent more tilt, whereas the long-dash line 5 per cent less tilt. The
horizontal level marks the ATHF extraction percentage. With the
exact power spectrum, the TOF is still able to detect 85 per cent of the point
sources if  the supplied $\theta$ is chosen in
between $\theta_{+}$ and $\theta_{-}$. } \label{comparison}
\end{figure}

We further have the following two assumptions. At the range of $k \la
\kmin$, the power spectrum of the combined signal $C(\kmin)$ is mainly determined by the
sky signals, and the pixel noise contribution
is not significant, so 
\begin{equation}
C(\kmin) \simeq C^{\rm sky}(\kmin). \label{eq:approkmin} 
\end{equation}
On the other hand, for the range of $k \ga \kmax$, the power spectrum is mainly
determined by the pixel noise power spectrum, which leads 
\begin{equation}
C(\kmax)\simeq \frac{C^{\rm pix}}{W^2(\kmax)}. \label{eq:approkmax}
\end{equation}
The assumptions of Eq.~(\ref{eq:asum}), (\ref{eq:approkmin}) and
(\ref{eq:approkmax}) are illustrated in Fig.~\ref{powerspectrum}, in which we
show the simulated power spectrum for the \planck High Frequency Instrument
(HFI) 100 GHz frequency channel. Eq.~(\ref{eq:approkmin}) and Eq.~(\ref{eq:approkmax}) lead to the
following relationship between $\kmax$ and $\kmin$ parameter,
\begin{equation}
\overline{B}(\kmax) \overline{B}(\kmin) = \frac{C^{\rm pix}}{C^{\rm
sky}(\kmin)},
\label{eq:equiv1}
\end{equation}
which leads to 
\begin{equation}
\kmax^2  \simeq  \frac{2}{\theta^2_B} \ln \left[\frac{C^{\rm sky}(\kmin)}{C^{\rm
pix}}\right]. 
\end{equation}
The ratio $R$ can then be obtained from Eq.(\ref{eq:G1}) 
\begin{equation}
R^{\rm ATHF}\simeq g^2\sum_i S_i^2 \frac{2}{\theta^2_B C^{\rm sky}(\kmin)}.
\end{equation}
We can further solve Eq.~(\ref{eq:G4}) by inserting Eq.~(\ref{eq:ctot}), which
then becomes (with the integral part of $C^{\rm pix} \ll C^{\rm sky} W^2$)
\begin{equation}
\int^{\kmax}_{\kmin} dk \;k \; C^{\rm sky}(k) W^2(k) \simeq \frac{ C^{\rm
sky}(\kmin)}{2 \theta^2}.  \label{approx}
\end{equation}
If the $C^{\rm sky}(k)=A k^{-n}$, Eq.~(\ref{approx}) can be approximated and we
reach
\begin{equation}
\kmax \sim \kmin + \frac{1}{\kmin \theta^2}.
\end{equation} 
Note that we use the simple trapezoidal rule for the integration, which works
better when the index $n$ is smaller.

\begin{figure}
\epsfig{file=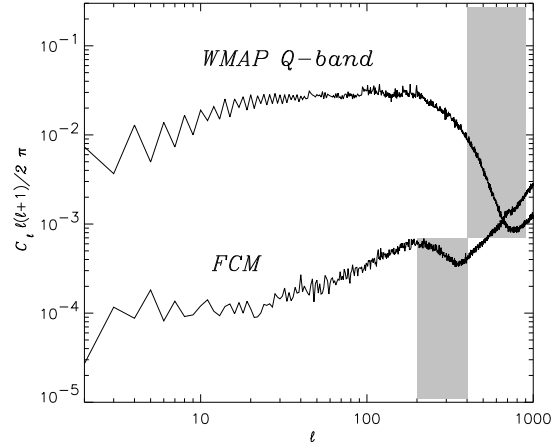,width=7.5cm}
\caption{The power spectrum of the foreground-cleaned map by Tegmark,
de Oliveira-Costa \& Hamilton (2003) and the \wmap Q-band map. The
shaded areas are the filtering ranges of the ATHF for point source
detection: $(\lmin,\lmax)=(200,400)$ for the FCM and $(400,900)$ for
the \wmap Q-band map.} \label{filteringrange}
\end{figure}

\begin{figure*}
\epsfig{file=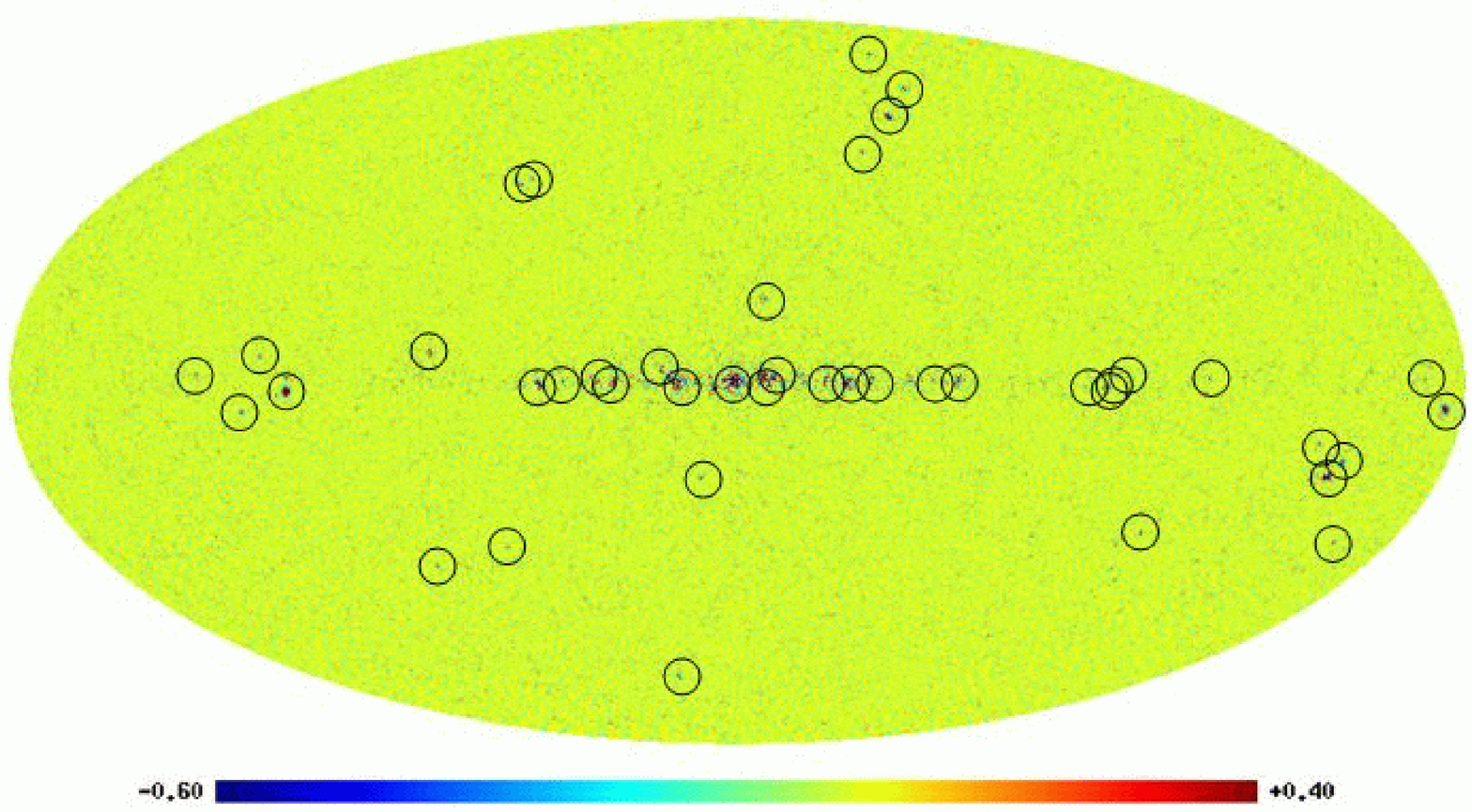,width=10cm,angle=-90}
\epsfig{file=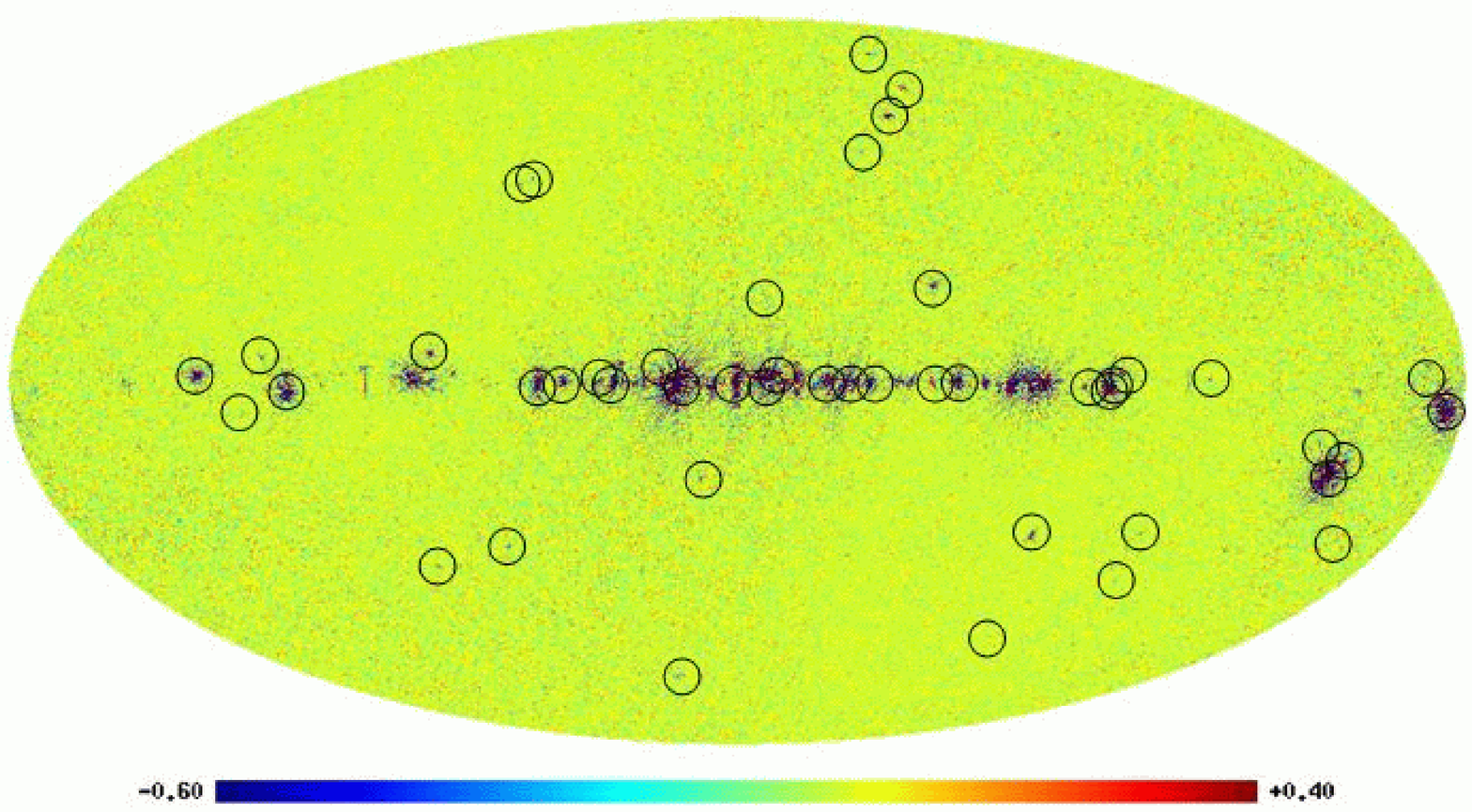,width=10cm,angle=-90}
\caption{The adaptive top-hat filtered maps from the foreground-cleaned map
(top) by Tegmark, de Oliveira-Costa \& Hamilton (2003) and the \wmap
Q-band map (bottom). The filtering ranges for both maps are shown in
Fig.~\ref{filteringrange}. In the top panel, the circled are peaks
with amplitudes above $5\sigma_{\rm f}$ after filtering, which are the
point source residues from the component separation. One interesting
feature is that not all the residues we recover are extragalactic point
sources. The dash circles are not peaks, but dips. In the bottom panel
we retain the positions of the circles for comparison, from which
those peaks are point sources. This is to
prove that most of the peaks we retrieve from the FCM are point source
residues.} \label{pointsource}
\end{figure*}

\section{Application of the ATHF on simulations and WMAP data}
\subsection{Simulations of the \planck HFI 100 GHz channel} 
We carry out the simulations of the \planck HFI 100 GHz 
channel for point source extraction by the TOF and the ATHF. The
simulation area is 25.6 ${\rm deg}^2$, and the pixel size is 3.0
arcmin. The $\sigma_{\rm CMB}$ and the $\sigma_{\rm pix}$ are $4.03
\times 10^{-5}$ and $6.07 \times 10^{-6}$, respectively. Point sources with amplitudes above
$2\sigma_{\rm CMB}$ at this channel would be flushed out by both (symmetric) filters with
the 5 $\sigma$ criterion (see the 5th column of Table 3 in Chiang
\etal \shortcite{tophat}), so we add randomly 50 point sources whose amplitudes
are $1.5 \sigma_{\rm CMB}$ to check the efficiency. Obviously the number of
point sources is too large for this frequency channel, but it serves as
the demonstration of the efficiency of extraction. We specifically use an
elliptical beam shape in the simulation with ellipticity ratio
$\theta_{+}/\theta_{-}=1.3$
\cite{burigana1998}. We designate the elliptical beam size such that
$\theta_{-}^2+\theta_{+}^2 =  2 \theta_0^2$ where
$\theta_0={\rm FWHM}/\sqrt{8 \ln 2}$. For the HFI 100 GHz channel the FWHM is 10.7
arcmin. In order to simulate more realistic situations, we
put this beam in rotation while scanning across the simulation area in order to
investigate how the elliptical beam could affect the efficiency of point source extraction
for the TOF and the ATHF.  

The period of rotation of the beam is $2 \pi$ along both sides of the simulation
square. Although the rotation period is too large for such size of map, it
nevertheless can elucidate the effect of elliptical beam shape regarding point
source extraction. 

Without the exact information of the in-flight beam size and orientation, the
optimal form of the TOF is by inserting a circular beam function into
Eq.~(\ref{eq:ro}), where the window function $W^2_k=\overline B_k^2 = \exp(-k^2
\theta^2)$. In Fig.~\ref{comparison} we plot the point source extraction rate
from the TOF for the 50 point sources. 
The plot is the extraction percentage of point sources
against the $\theta$ inserted to the TOF (in terms of 100 GHz channel $\theta_0$). We would
like to examine, for in-flight elliptical beam shapes, how the supplied beam
function for the TOF
can affect the extraction rate. In the dotted curve, the power spectrum is
precisely known. The short-dash curve is the case when the slope of the CMB
power spectrum has 5 per cent more tilt, whereas the long-dash line 5 per cent
less tilt. The horizontal line marks the ATHF extraction percentage. With the
exact power spectrum, the TOF is able to detect 85 per cent of the point sources
with amplitude $1.5 \sigma_{\rm CMB}$ when the supplied $\theta$ is chosen in between
$\theta_{+}$ and $\theta_{-}$. For the case of 5 per cent more tilt in power spectrum,
the supplied beam function will need adjustment to a smaller $\theta$ value to
ensure the TOF to have a maximal resonance, and adjustment to a larger $\theta$
for the case of 5 per cent less tilt. On the other hand, the ATHF can reach 74
per cent of extraction rate without any information. From Fig.~\ref{comparison}
it is not surprising that the 'safe bet' of the beam function is
$\theta/\theta_0 \simeq 1$ for the uncertainties in the power
spectrum. However, as the cleaning of foreground contamination precedes the
determination of the angular power spectrum, and during the whole sky scan of
the \planck mission, the possible degradation effect of the mirrors
could change the beam size and shape, the efficiency of the TOF will
be hampered.  

\subsection{The ATHF on the WMAP derived map} 
In this subsection we apply the ATHF to the derived map from the \wmap 1-year
data\footnote{http://lambda.gsfc.nasa.gov/product/wmap/}. The derived maps are
produced by Tegmark, de
Oliveira-Costa \& Hamilton \shortcite{tegwmap} performing an independent
component separation analysis
from the \wmap team. They are the foreground-cleaned map (FCM) and
the Wiener-filtered map (WFM). The FCM was tested to be
non-Gaussian \cite{wmapng}, partly due to galactic emission. Here we
investigate the possible point source residues from their component
separation. 

We use the ATHF on the FCM as it is the map with scientific significance. In
Fig.~\ref{filteringrange} we show the power spectra of
the FCM and the \wmap Q-band map with the corresponding filtering
ranges (the shaded area) of the ATHF: $(\lmin,\lmax)=(200,400)$ for the FCM and
$(400,900)$ for the \wmap Q-band map. It is
easy to see that the CMB (plus foreground) power spectra (convolved with the
beam) and the noise level. Note that the presentation of the angular power
spectrum in Fig.~\ref{filteringrange} is the standard format, i.e. $\Cl$ with the factor
$\ell(\ell+1)/4 \pi$, which is different from Fig.~\ref{powerspectrum} used for
analysis in the previous Section. We therefore apply the ATHF with 
filtering ranges covering the conjunction of the CMB and noise power curves. The choice of the filtering
range follows the rule of the thumb described in Chiang et al.
\shortcite{tophat}: the $\lmin$ to exclude most of the CMB power, the $\lmax$
the noise power, and both to keep the part being convolved by the beam. 

Fig.~\ref{pointsource} shows the filtered maps of the FCM (top panel) and the
\wmap Q-band map (bottom). The circled in the top panel are filtered peaks 
with amplitudes above $5\sigma_{\rm f}$. One interesting feature in
Fig.~\ref{pointsource} is that not all
the residues we recover are extragalactic point sources. The dash circles are
not peaks, but dips. In the bottom panel we retain the positions of
the circles for comparison, from which those peaks are point
sources. This is to prove that most of the peaks we retrieve from the
FCM are point source residues. We do not intend to recover all
residues from one single iteration of the ATHF but simply demonstrate
the usefulness and quickness of the ATHF on the real data. Application
of the ATHF on the \wmap 5  different channel maps will be on a
separate paper.

\begin{figure}
\centering
\epsfig{file=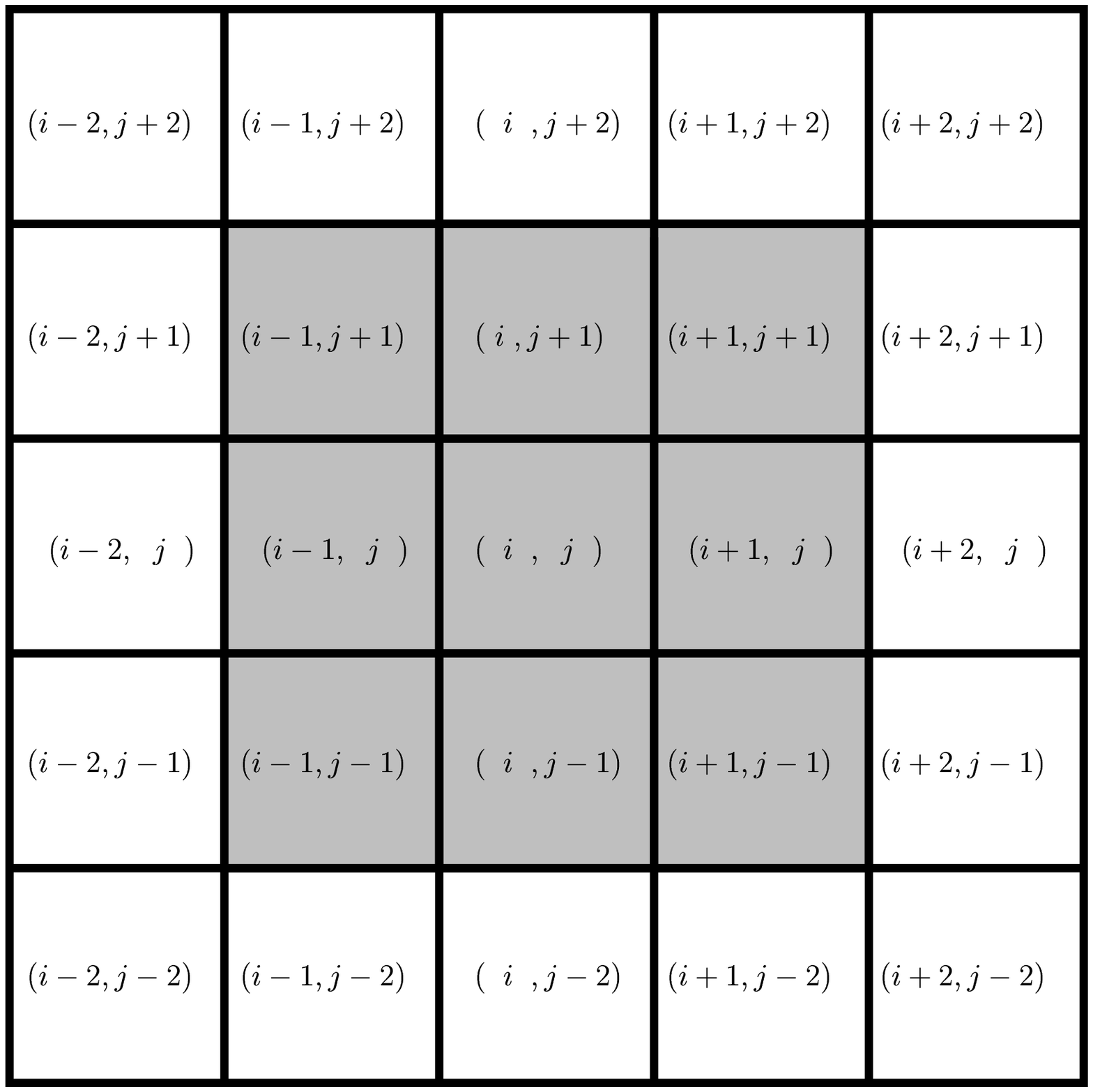,width=7.5cm}
\caption{The cascading hybrid median filtering scheme. For a residual
peak located at $(i,j)$ from the cleaning by circular Gaussian
profile, we can apply the cascading hybrid median filter to replace
the inner $3 \times 3$ pixels. First of all, we replace the central
pixel $(i,j)$ by the median of the 8 pixels from the same row, column,
and the diagonals outside the targeted region, i.e., by the median of pixels
$(i-2,j)$, $(i-2,j+2)$, $(i,j+2)$, $(i+2,j+2)$, $(i+2,j)$,
$(i+2,j-2)$, $(i,j-2)$ and $(i-2,j-2)$. Once the central pixel is
decided, we can in turn replace the four corners by the median of its
diagonals, e.g., the pixel $(i+1,j+1)$ can be replaced by the median
of $(i,j+2)$, $(i+2,j+2)$, $(i+2,j)$ and the replaced $(i,j)$. Then
the last 4 pixels can be replaced by their neighbouring pixels of the
same row and column, e.g., $(i+1,j)$ by the median of $(i+1,j+1)$,
$(i+2,j)$, $(i+1,j-1)$ and the replaced $(i,j)$.} \label{hybridmedian}
\end{figure}

One important issue related to elliptical beam-convolved point source
extraction is the removal of point sources. Normally, circular Gaussian
profile is used for cleaning the elliptical beam-convolved point sources from
the map after detection of the point source position. Due to the ellipticity of
the beam and pixel noise, there are considerable residues after the cleaning by
circular Gaussian profile. To eliminate the residual peaks, we apply a simple
algorithm that is modified from the so-called hybrid median filter used in
signal processing. The hybrid median filter is useful in noise reduction and
in particular in elimination of shot noise. This non-linear filter
replaces the targeting pixel with average of the neighboring pixels from
the same row, column and two diagonals. In the modified cascading
version, the targeted region ($ 3 \times 3$ pixels) centered
at the residual peak is replaced with the following algorithm (see
Fig.~\ref{hybridmedian}). The central pixel is replaced by the median
of the 8 pixels from the same row, column, and two diagonals right
outside the targeted region. The four pixels at the four corners in
the region can then be decided by the median of their 4 corner pixels,
which include the already-replaced central pixel. The rest 4 pixels can be
replaced by the median of their neighbouring pixels from the same row
and column.

\section{Conclusions}
We have discussed analytically the three main linear filters for point source
extraction of the \planck mission. The TOF is optimal in terms of the gain
factor. The so-called optimal pseudo-filter is at its best asymptotic at the far
ends to the TOF under certain circumstance. Both of these filters require the
experiment parameters such as the beam shape and size.
 
Note that the calculation is based on the simple geometrical model of
the beam. As we mention in the beginning, the point source is the
reflection of, among others, the beam shape properties. The possible antenna    
beam degradation would be a problem for the calibration of the
in-flight antenna beam shape \cite{degradation}, causing the change of the
inclination and the ellipticity ratio of the beam during the mission.

There is one comment on the extraction of point sources from the
time-ordered scans when the inclination of the beam, the issue of pointing and 
the noise properties are most crucial: the inclination angle and the pointing
will decide the point source FWHM on the time-ordered data, for which the ATHF
is particularly useful.   

Note also that we can easily generalize the multi-frequency method for point
source extraction \cite{multifreq} by applying the ATHF for all
\planck frequency channels. 

\section*{Acknowledgments}
This paper was supported by Danmarks Grundforskningsfond
through its support for the establishment of the Theoretical
Astrophysics Center. The authors are grateful for
Tegmark et al. for their processed maps. We acknowledge the use of
{\sc healpix} \cite{healpix} package and the {\sc glesp} code
\cite{glesp} for whole-sky data processing. We thank O. Verkhodanov
for the help with the {\sc glesp} package.

\end{document}